\documentclass[notitlepage,10pt,aps,prd,twocolumn,amsmath,amssymb,floatfix,superscriptaddress,nofootinbib,preprintnumbers,groupedaddress]{revtex4-2}

\usepackage{amsfonts,amssymb,amsmath,graphicx,color,bm,epsfig,dsfont}

\def\Mpc{{\rm Mpc}}
\def\Msun{M_\odot}
\def\Mpl{M_{\rm Pl}}

\def\keq{k_{\rm eq}}
\def\Meq{M_{\rm eq}}
\def\aeq{a_{\rm eq}}

\def\kR{k_R}

\def\PBH{{\rm PBH}}
\def\fPBH{f_\PBH}

\def\fMD{f_{\rm MD}}

\def\itime{0}

\def\kmin{k_{\rm min}}

\def\aH{a_{H}}
\def\aHmatch{\tilde{a}_{H}}
\def\aNL{a_{\rm NL}}
\def\Hmatch{\tilde{H}}

\def\sigmaHC{\tilde{\sigma}(R, \aHmatch) }
\def\deltaHC{\tilde{\delta}(R, \aHmatch) }

\DeclareMathOperator\erf{erf}
\DeclareMathOperator\erfc{erfc}

\definecolor{ultramarine}{rgb}{0.07, 0.04, 0.56}
\definecolor{cadmiumgreen}{rgb}{0.0, 0.42, 0.24}
\definecolor{indigo(dye)}{rgb}{0.0, 0.25, 0.42}
\usepackage[linktocpage=true, bookmarks=false]{hyperref}
\hypersetup{
colorlinks=true,
citecolor=ultramarine,
linkcolor=cadmiumgreen,
urlcolor=indigo(dye),
}

\newcommand\numberthis{\addtocounter{equation}{1}\tag{\theequation}}

\definecolor{darkgreen}{cmyk}{0.85,\itime.2,1.00,\itime.2} 
\definecolor{purple}{cmyk}{0.5,1.0,\itime,\itime} 

\begin{document}

\preprint{YITP-21-94}

\title[]{Primordial Black Holes from CDM Isocurvature Perturbations}

\label{firstpage}

\author{Samuel Passaglia$^a$}
\email{samuel.passaglia_at_ipmu.jp}
\author{Misao Sasaki$^{a,b,c}$}
\email{misao.sasaki_at_ipmu.jp}
\affiliation{$^a$Kavli Institute for the Physics and Mathematics of the Universe (WPI),The University of Tokyo Institutes for Advanced Study (UTIAS),The University of Tokyo, Chiba 277-8583, Japan}
\affiliation{$^b$Center for Gravitational Physics, Yukawa Institute for Theoretical Physics, Kyoto University, Kyoto 606-8502, Japan}
\affiliation{$^c$Leung Center for Cosmology and Particle Astrophysics, National Taiwan University, Taipei 10617, Taiwan}

\begin{abstract}
We show that primordial black holes can be produced from the collapse of large isocurvature perturbations of the cold dark matter. We develop a novel procedure to compute the resulting black hole abundance by studying matched perturbations of matter-only universes, and we use our procedure to translate observational constraints on black hole abundances into model-independent constraints on cold dark matter isocurvature perturbations over a wide range of scales. The constraint on the typical amplitude of the primordial perturbations weakens slightly slower than linearly on small scales.
\end{abstract}

\date{\today}
\maketitle 

\section{Introduction}
\label{sec:intro}

	Our universe was endowed on large scales with primordial fluctuations which are small and adiabatic, visible directly in the anisotropy pattern of the Cosmic Microwave Background \cite{COBE:1992syq,Akrami:2018odb}. We have much less information about the initial state of our universe on small scales, where primordial information is processed by nonlinear physics.

	One model independent statement that can be made about small scale perturbations is that they cannot be so large as to form too many primordial black holes (PBHs). PBHs form from the collapse of large adiabatic perturbations when they enter the cosmological horizon \cite{Zeldovich:1967aaa,Carr:1974,Carr:1975aaa}, and therefore astronomical measurements of black hole abundances provide invaluable cosmological information about the primordial adiabatic mode on small scales \cite{Carr:1993aq,Gow:2020bzo}.

	In this work, we show that PBHs can also form from the collapse of primordial {\it isocurvature} fluctuations of the cold dark matter (CDM). This allows us to use black hole abundance data to place wholly model-independent constraints on the CDM isocurvature component of the small scale perturbations.

	We will see that CDM isocurvature perturbations can form PBHs if they are so large that the local matter density becomes nonlinear sufficiently close to the horizon scale. Forming light PBHs, corresponding to small regions entering the horizon early during radiation domination, therefore requires CDM perturbations large enough that a local region becomes CDM dominated well before global matter-radiation equality. Constraints on primordial CDM isocurvature from PBHs therefore become weaker and weaker on smaller and smaller scales as larger and larger CDM perturbations are required to induce local matter domination earlier and earlier. This is in sharp contrast with the adiabatic case, where the perturbation amplitude required to produce a given PBH abundance at formation does not significantly depend on when the PBHs form.	

	The exact formation probability which relates the amplitude of CDM isocurvature perturbations to the PBH abundance has not yet been computed in the literature. In this work we develop a novel matching procedure to estimate it by identifying CDM isocurvature modes with perturbations in matched matter-only universes. This matching allows us to adapt results from studies of PBH formation in matter-dominated universes \cite{Harada:2016mhb,Harada:2017fjm,Kokubu:2018fxy} to study PBH formation from CDM isocurvature. Because PBH formation in matter domination is only power-law sensitive to the perturbation amplitude, rather than exponentially sensitive as in the adiabatic radiation-domination case, constraints on PBH abundances probe a greater dynamic range of the isocurvature spectrum than the adiabatic spectrum.  

	This paper is organized as follows. 

	In \S\ref{sec:evolution}, we recall some basic results about the linear evolution of CDM isocurvature perturbations in the early universe.

	In \S\ref{sec:collapse}, we describe how sufficiently large CDM isocurvature fluctuations can collapse to form PBHs and we introduce our matching scheme to compute the PBH abundance today from a given amplitude of primordial isocurvature fluctuations.

	In \S\ref{sec:constraints}, we apply our technique to map existing PBH abundance constraints to constraints on the primordial CDM isocurvature. 

	Finally in \S\ref{sec:discussion} we discuss and contextualize our results.

\section{Evolution of CDM Isocurvature Perturbations}
\label{sec:evolution}

	In this section we wish to solve for the evolution of a CDM perturbation {$\delta_c \equiv \delta \rho_c  / \rho_c$} starting from a large isocurvature initial condition. By this we mean that we formally decompose the CDM perturbation at the primordial epoch into isocurvature and adiabatic components \cite{Bucher:1999re} 
	\begin{equation}
	\delta_c(0) = S(0) + A(0),
	\end{equation}
    as
    \begin{equation}
    \begin{split}
	S(0) &= \delta_c(0) - 3/4 \delta_\gamma(0), \\
	A(0) &= 3/4 \delta_\gamma(0),
	\end{split}
	\end{equation}
	with $\delta_\gamma \equiv \delta \rho_\gamma / \rho_\gamma$ the photon overdensity. $(0)$ denotes an initial time deep in radiation domination when all scales of interest are superhorizon. We restrict our attention to the large isocurvature regime $S(0) \gg A(0)$.

	We work throughout in the Newtonian gauge, where the metric takes the form
	\begin{align}
	ds^2=a^2(\tau)\left[-(1+2\Psi)d\tau^2+(1+2\Phi) \delta_{ij} dx^idx^j\right],
	\end{align}
	with $\Psi$ and $\Phi$ the Newtonian gauge curvature perturbations. These are controlled in radiation domination by the radiation perturbations through the Einstein equations \cite{Kodama:1986fg,Kodama:1986ud,Ma:1995ey}, and so the large isocurvature regime corresponds to $\delta_c \gg \left\{\Phi, \Psi\right\}$ during radiation domination for as long as the total density and metric perturbations remain linear.

	We can therefore neglect the metric perturbations in the continuity and Euler equations for the CDM in linear theory in radiation domination. We also restrict our attention to modes which enter the horizon in radiation domination, and which are therefore subhorizon during matter domination. Since metric perturbations decay on subhorizon scales, we can therefore neglect them in the continuity and Euler equations for the CDM in matter domination as well. Under these approximations the CDM perturbations evolve as \cite{Hu:1995en}
	\begin{equation}
	\label{eq:CDMevol}
	\frac{d^2 \delta_c}{dy^2} + \frac{(2+3y)}{2 y ( 1+y)} \frac{d\delta_c}{d y}  = \frac{3}{2 y(1+y)} \frac{\Omega_{c}}{\Omega_{m}} \delta_c,
	\end{equation}
	where $y\equiv a/\aeq$ is the scale factor relative to matter-radiation equality $\aeq$ and $\Omega_c$ and $\Omega_m$ are, respectively, the CDM and total matter densities today. For simplicity, we throughout this work neglect the energy contribution of the baryons and therefore  set $\Omega_{c} = \Omega_{m}$. 

	The solutions to Eq.~\eqref{eq:CDMevol} are
	\begin{align*}
	U_1 &= \frac{2}{3} + y \numberthis, \\ 
	U_2 &=  \frac{15}{8} (2 + 3 y) \ln\left[ \frac{(1+y)^{1/2} + 1}{(1+y)^{1/2} - 1} \right] - \frac{45}{4} (1 +y)^{1/2},
	\end{align*}
	of which only $U_1$ approaches a constant during radiation domination $y\rightarrow0$. Matching to this solution, we conclude that the CDM perturbation evolves as
	\begin{equation}
	\delta_c \simeq \left(1+ \frac{3}{2} \frac{a}{\aeq}\right) S(0),
	\end{equation}
	which defines the transfer function
	\begin{equation}
	T(a) \equiv \frac{\delta_c(a)}{\delta_c(0)} \simeq 1 + \frac{3}{2}\frac{a}{\aeq}.
	\end{equation}
	This transfer function evolves CDM fluctuations from their initial value, through radiation domination and into matter domination, until linear theory breaks down as signaled by the total density perturbation becoming order unity.
	
	Defining $\keq$ to be the mode which crosses the horizon at matter-radiation equality,
	\begin{equation}
	\keq = \left(2 \Omega_{m} H_0^2 / \aeq\right)^{1/2} \simeq 0.01 \Mpc^{-1}, 
	\end{equation}
	where $H_0$ is the Hubble constant today, this transfer function is valid for $k\gg \keq$. These modes enter the horizon at $a_H$,
	\begin{align*}
	\label{eq:aH}
	\frac{\aH}{\aeq} = \frac{1 + \sqrt{1+8\left(k/\keq\right)^2}}{4(k/\keq)^2}
	&\simeq \frac{1}{\sqrt{2}}\frac{\keq}{k} \quad (k \gg \keq) \numberthis,
	\end{align*}
	well before matter radiation equality $\aeq\gg \aH$.
	
\section{PBHs from CDM collapse}
\label{sec:collapse}

	The CDM overdensity can be spatially averaged across a comoving scale $R$ as
	\begin{equation}
	\delta_c(R) = \int \frac{d^3 \vec{k}}{2 \pi^3} \  W_R(k) \delta_c(\vec{k}),
	\end{equation}
	where $W_R$ is a smoothing function which suppresses small-scale $k \gg R^{-1}$ modes and $\delta_c(\vec{k})$ denotes a Fourier mode of the CDM density perturbation. The time dependence we solved for in \S\ref{sec:evolution} can be extracted from the integral to write
	\begin{equation}
	\delta_c(R) = T(a) \delta_c(R, \itime),
	\end{equation}
	with $\delta_c(R, \itime)$ encoding the primordial isocurvature $S(0)$ smoothed by $W_R$.

	The local CDM mass contained within a physical scale $a R$ depends on the smoothing function chosen, and we absorb this dependence into a parameter $\mu$, 
	\begin{align*}
	\label{eq:MtoR}
	M(R) &= \mu \times \frac{4}{3} \pi (a R)^3 \times \rho_{c}(R)
	\\&=  4 \pi R^3 \mu \Omega_{c} \left(1+\delta_c(R, \itime)\right) H_0^2 \Mpl^2. \numberthis
	\end{align*}
	The window function also associates with $R$ a comoving wavenumber $\kR \equiv \gamma/R$, with $\gamma$ some number. Though our results are qualitatively insensitive to such choices, in our examples we use $\mu = (9 \pi / 2)^{-1}$ and $\gamma = 2.7$ which correspond to a Fourier space top-hat window function up to some assumptions about convergence of the volume integral \cite{Schneider:2013ria,Young:2019osy}.

    Whether a region collapses into a PBH depends on the local configuration of matter, and the PBH abundance can therefore be obtained by integrating the probability distribution from which the perturbations are drawn against some PBH formation criterion. The PBH abundance at formation is defined as
	\begin{equation}
	\beta(R) \equiv  \left.\frac{1}{\rho_{\rm tot}} \frac{d \rho_\PBH(R)}{d \ln R} \right\vert_{\rm formation},
	\end{equation}
	which is the differential energy density contained in PBHs formed by perturbations on the scale $R$, relative to the total energy density and evaluated at the formation time. 

    For adiabatic fluctuations, the PBH formation criterion and the resulting PBH abundance have been studied in detail. If the smoothed total density fluctuations $\delta(R)$ are drawn from a Gaussian with variance $\sigma^2(R)$ then
	the abundance of PBHs at formation can be computed as some function $f$ of the typical fluctuation amplitude at horizon crossing $\sigma(R,\aH)$,
	\begin{equation}
	\beta(R) = f(\sigma(R, \aH)) \qquad \text{(adiabatic)}.
	\end{equation}

	For adiabatic perturbations in radiation domination (see, e.g., Ref.~\cite{Motohashi:2017kbs}), $f = f_{\rm RD}$ is an error function counting regions in the tail of the density distribution which lie over a collapse threshold where the radiation pressure can be overcome. 

	For adiabatic perturbations in matter domination (see, e.g., Ref.~\cite{Ballesteros:2019hus}), the fluid is pressureless and for large fluctuations $f = \fMD$ is only power law suppressed, with Ref.~\cite{Harada:2016mhb,Harada:2017fjm,Kokubu:2018fxy} deriving
	\begin{align*}
	\label{eq:fharada}
	\fMD(\sigma(R, \aH)) &\simeq \fMD^{\rm inhomogeneous}\times \fMD^{\rm anisotropic} \\ &=  0.02  \ \sigma(R, \aH)^{13/2},\numberthis
	\end{align*} 
	valid for $0.005 \lesssim \sigma(R, \aH) \lesssim 0.1$. A slightly different behavior holds above $\sim0.1$, while below $\sim0.005$  the PBH production is exponentially suppressed. The exact form of the exponential cutoff including all relevant physics is under investigation, but as a preliminary estimate the results of Refs.~\cite{Harada:2017fjm,Kokubu:2018fxy} can be combined as
	\begin{align*}
	\label{eq:fharada2}
	\numberthis \fMD(\sigma(R,\aH)\lesssim&\ 0.005) \simeq\\ 4 &\times 10^{-7} \sigma(R,\aH)^{7/2} e^{-0.15 \sigma(R,\aH)^{-2/3}},
	\end{align*}
	where the coefficient has been chosen to match Eq.~\eqref{eq:fharada} at the switch point. 

	In contrast to the adiabatic case, the formation of PBHs from primordial CDM isocurvature fluctuations has not yet been examined in the literature. In particular the mapping between the amplitude of the CDM isocurvature fluctuations and the resulting PBH abundance is not yet known.

	Our novel approach is to estimate the PBH abundance by matching the CDM isocurvature fluctuation $\delta_c (R)$ to a fluctuation $\tilde{\delta}(R)$ of a matter-only universe. 

	Namely, we construct a fictitious CDM-only universe which behaves identically to the actual universe once the CDM has dominated the local energy density. We insert into this universe a perturbation $\tilde{\delta}(R)$ which matches in amplitude and scale the total density perturbation $\delta (R)$ in our universe. Tildes denote quantities in the fictitious universe here and throughout.

	The matched perturbation $\tilde{\delta}(R)$ can then be evolved back in time to find its amplitude $\deltaHC$ when it crossed the horizon in the CDM-only universe to determine whether it, and therefore the isocurvature fluctuation it matches, will form a PBH. This matching allows us to identify the PBH abundance produced by Gaussian isocurvature fluctuations with typical initial amplitude $\sigma_c(R, \itime)$ with the PBH abundance produced by the matched $\sigmaHC$ as $\fMD(\sigmaHC)$ using the matter-only formulas Eq.~\eqref{eq:fharada} and Eq.~\eqref{eq:fharada2}. 

	To perform the matching we define a time $\aNL(R)$ when the total overdensity smoothed on the scale $R$, 	
	\begin{equation}
	\delta(R) = \frac{\rho_c}{\rho_c+\rho_R}  \delta_c(R) \simeq \delta_c(R, \itime) \frac{(1+ 3 a / 2 \aeq)}{\left(1 + \aeq/a\right)},
	\end{equation}
	is equal to $1$,
	\begin{equation}
	\delta(R, \aNL) \equiv 1,
	\end{equation}
	and we suppress the $R$ dependence of $\aNL(R)$ here and throughout. 
	Explicitly, we have
	\begin{equation}
	\label{eq:aNL}
	\frac{\aNL}{\aeq} =  \frac{B}{\delta_c(R, \itime)}, 
	\end{equation}
	where $B = (1 - \delta_c(R,\itime) + \sqrt{1 + 4 \delta_c(R,\itime) + \delta_c(R,\itime)^2})/3$ is an order unity number that goes from $2/3$ when $\delta_c(R,\itime) \ll 1$ to $1$ when $\delta_c(R,\itime)\gg1$.

	$\aNL$ indicates when the evolution of the CDM perturbation becomes nonlinear.	We identify two limits of its behavior. 

	When $\delta_c(R, \itime)$ is less than unity, we have $\aNL>\aeq$ and nonlinearity begins during the globally matter dominated period of our universe. In this regime, the CDM fluctuation is frozen until matter domination begins, and then grows $\propto a$ until linear theory breaks down at $\aNL$.

	When $\delta_c(R, \itime)$ is greater than unity, on the other hand, we have $\aNL < \aeq$ and nonlinearity begins during global radiation domination while $\delta_c(R)$ is constant. At $\aNL$, local matter domination begins and $\delta_c(R)$ begins to evolve nonlinearly. Comparing $\aNL$ to the horizon crossing time $\aH$,
	\begin{equation}
	\label{eq:aNLoveraH}
	\frac{\aNL}{\aH} \simeq \frac{ \sqrt{2} }{\delta_c(R, \itime)} \frac{\kR}{\keq},
	\end{equation}
	we restrict our attention to perturbations $\delta_c(R,\itime) \lesssim \sqrt{2} \kR/\keq$ so that nonlinearity begins only once the perturbation is subhorizon.

	\begin{figure}[t]
	\includegraphics{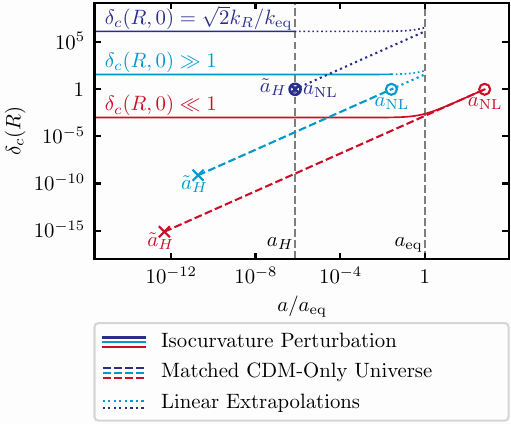}
	\caption{The matching procedure which associates CDM isocurvature modes with fluctuations of matter-only universes. We show CDM fluctuations $\delta_c$ on a fixed scale $\kR/\keq = 10^6$ ($R\simeq \gamma \times 10^{-4} \ \Mpc$) crossing the horizon at $\aH/\aeq \simeq 10^{-6}$. A small fluctuation (solid red) is constant in radiation domination and grows in matter domination until the total density fluctuation $\delta \simeq \delta_c \times \rho_c/\rho_{\rm tot}$ becomes nonlinear ($\aNL$, circle). We match it to a perturbation of a CDM-only universe (dashed red) with the same Hubble rate at $\aNL$ and a smaller amplitude $\deltaHC$ at horizon crossing in the matched universe ($\aHmatch$, cross). For a larger fluctuation (blue), nonlinearity occurs in radiation domination, the matched universe has a larger Hubble rate, and the matched perturbation a larger $\deltaHC$. For a large enough fluctuation (purple), $\aNL$, $\aH$, and $\aHmatch$ are simultaneous and collapse occurs at horizon crossing. Still larger fluctuations break our formalism but overproduce PBHs. Linear extrapolations (dotted) help show the amplitude of the fluctuations. See \S\ref{sec:collapse} for further discussion.}
	\label{fig:backtracking}
	\end{figure}

	We now use $\aNL$ to construct our fictitious matter-only universe by matching the Hubble rate between the true and fictitious universes at $\aNL=\tilde{a}_{\rm NL}$. The Friedmann equation in the fictitious universe reads
	\begin{align*}
	\Hmatch^2(\tilde{a}) &= H(\aNL)^2 \left(\frac{\tilde{a}}{\aNL}\right)^{-3} \\ &= \frac{H_0^2 \Omega_{m}}{\tilde{a}^3} \left( \frac{\aeq}{\aNL}+1\right), \numberthis
	\end{align*}
	and in this matter-only universe horizon crossing for the scale $R$ occurs at
	\begin{equation}
	\label{eq:amatch}
	\frac{\aHmatch}{\aeq} =  \frac{\keq^2}{2\kR^2} \left( \frac{\aeq}{\aNL}+1\right), 
	\end{equation}
	where recall we only consider small-scale perturbations $\kR \gg \keq$.

	We then insert into this fictitious universe a perturbation $\tilde\delta(R)$ with amplitude at horizon crossing 
	\begin{align*}
	\label{eq:deltaHC}
	\deltaHC &=  \frac{\aHmatch}{\aNL} \\&=  \frac{\keq^2}{2\kR^2} \frac{\delta_c(R,\itime)}{B} \left(\frac{\delta_c(R,\itime)}{B}+1\right), \numberthis
	\end{align*}
	which enters the horizon earlier and with a lower amplitude than the CDM isocurvature mode but its additional linear growth $\propto a$ rescales its amplitude such that at $\aNL$ it matches the density perturbation in our universe exactly, $\tilde{\delta}(R, \aNL) = \delta(R, \aNL) = 1$. We then identify PBH formation from the isocurvature fluctuation with PBH formation from this matching perturbation of the matter-only universe.

	For Gaussian CDM fluctuations with variance
    \begin{equation}
	\sigma_c^2 (R) = \int \frac{d k}{k} W_R(k)^2 \frac{k^3 \langle {\delta_c (\vec{k})^2} \rangle}{2 \pi^2},
	\end{equation}
	{the variance of the matched fluctuations is thus}
	\begin{equation}
	\label{eq:sigmaHC}
	\sigmaHC^2 =  \frac{\keq^4}{4\kR^4} \frac{\sigma_c(R,\itime)^2}{B^2} \left(3 \frac{\sigma_c(R,\itime)^2}{B^2}+1\right),
	\end{equation}
	with $B \sim B(\sqrt{\sigma_c(R,\itime))^2})$.

	From this matched amplitude of fluctuations the PBH abundance at formation is computed using the matter-only formula as
	\begin{equation}
	\label{eq:beta}
	\beta(R) = \fMD(\sigmaHC),
	\end{equation}
	and therefore we can compute the PBH abundance for CDM isocurvature modes by adapting existing results for PBH formation from adiabatic perturbations in matter-domination.

	Note that the matching parameters $\deltaHC$ and $\sigmaHC$, which control the PBH abundance, are suppressed for small-scale modes by a factor $(\keq/\kR)^2$. This occurs because the Newtonian potential is suppressed inside the horizon, and at $\aNL$ it is
	\begin{align*}
	\label{eq:potential}
	\Psi(R, \aNL) &\simeq \frac{\aNL^2 H(\aNL)^2}{\kR^2} \delta(R, \aNL)
	\\ &\simeq  \frac{\delta_c(R,\itime)^2}{2}  \frac{\keq^2}{\kR^2}, \numberthis
	\end{align*}
	where we have used that $\delta(R, \aNL) = 1$ by definition. With $\Psi(R, \aNL)$ suppressed by $(\keq/\kR)^2\ll1$, significant nonlinear growth has to occur after $\aNL$ for PBHs to form ($\Psi \sim 1$) unless $\delta_c(R,\itime)$ is very large. This nonlinear growth enhances anisotropies in the distribution and makes collapse to PBHs more difficult.

	\begin{figure}[t]
	\includegraphics{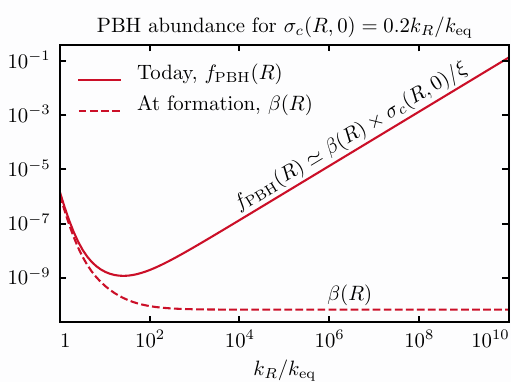}
	\caption{The primordial black hole abundance today (solid) and at formation (dashed) as a function of formation scale $\kR/\keq$ for Gaussian primordial isocurvature fluctuations with primordial amplitude $\sigma_c(R,\itime) = 0.2 \kR / \keq$. As $\kR$ increases, the  $(\kR/\keq)^{-2}$ suppression of $\sigmaHC$ \eqref{eq:sigmaHC} is compensated by the larger and larger input perturbation and the PBH abundance at formation approaches a constant. This large input perturbation leads the PBHs to form earlier in radiation domination, enhancing the abundance today through the growth function $g(R) \simeq \sigma_C(R,\itime)/\xi$ \eqref{eq:growth}. See \S\ref{sec:collapse} for further discussion.}
	\label{fig:betaforsigma}
	\end{figure}

	Therefore it is only when $\delta_c(R,0)$ is not much less than $\sqrt{2} (\kR/\keq)$ that PBHs can form, or equivalently only when $\aNL \sim \aH$ (see Eq.\eqref{eq:aNLoveraH}). In this regime the CDM perturbation is so large that local matter domination occurs near horizon crossing for the mode, deep in global radiation domination. When $\aNL = \aH$, the total smoothed density fluctuation of order unity at horizon crossing and subhorizon physics cannot prevent collapse. This represents the upper limit of validity of our calculation and corresponds to  the matched mode amplitude at horizon crossing $\deltaHC = 1$. 

	{In fact, Ref.~\cite{Domenech:2021and} found by exact solution that the curvature perturbation induced near horizon crossing by a primordial isocurvature perturbation $S(0) \sim \delta_c(0)$ can be qualitatively approximated during radiation domination by an effective primordial curvature perturbation with amplitude $\zeta(0) \sim (k_{\rm eq}/k_R) S(0)$. This further reinforces that PBH formation is possible when the isocurvature perturbation is large enough to balance the suppression factor.}

	{When this occurs our matching procedure then appropriately predicts large PBH abundances. For perturbations even larger than this our formalism breaks down, but such perturbations inevitably overproduce black holes.}
	
	\begin{figure*}[p!]
	\psfig{file=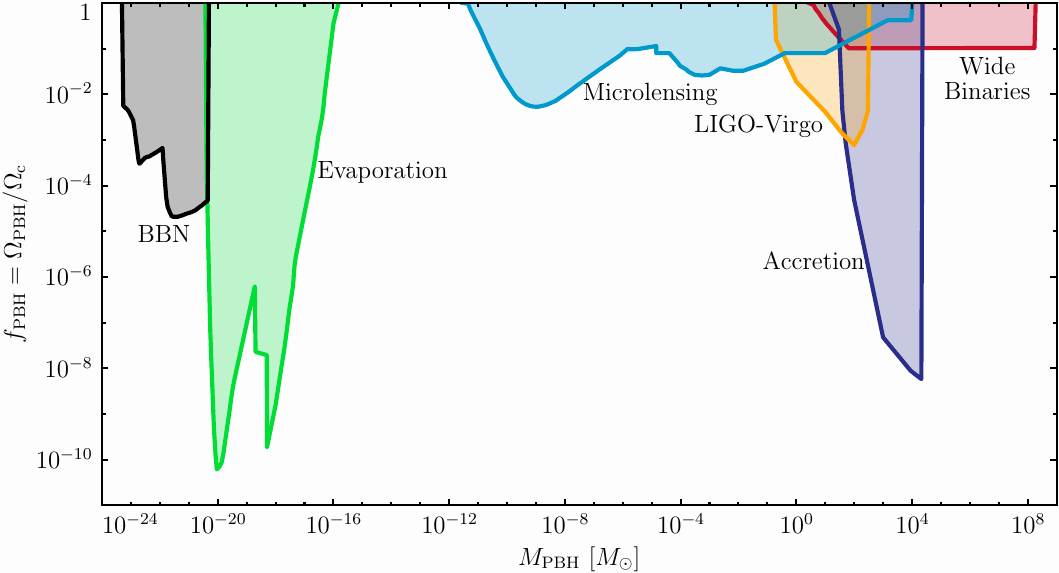}

	\vspace{.5cm}

	\psfig{file=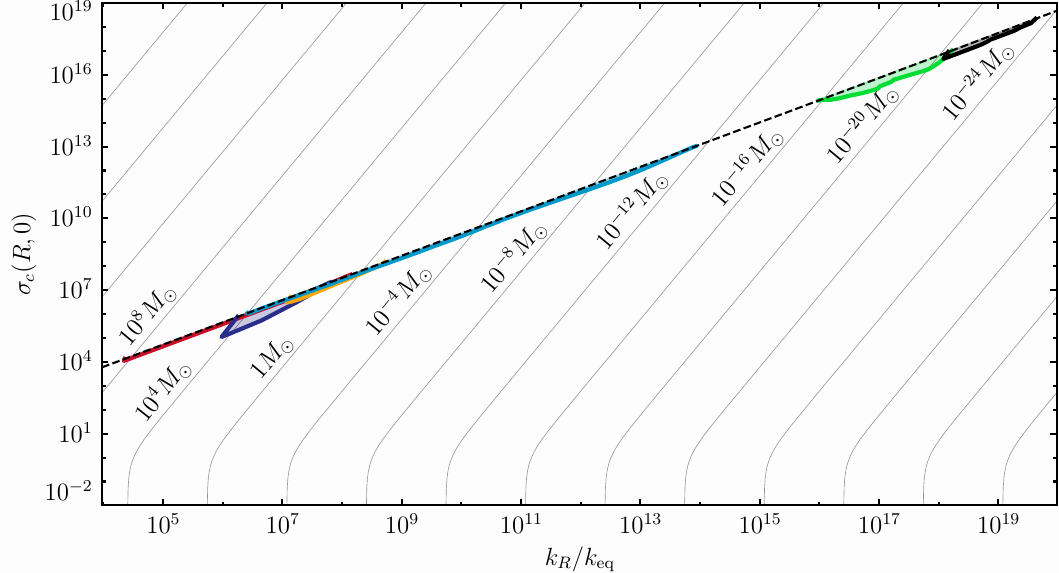}
	\caption{Observational constraints on the PBH abundance relative to the CDM $\fPBH$ (top) can be mapped to constraints on the primordial CDM overdensity $\sigma_c(R,\itime)$ (bottom). The scale $\kR=\gamma/R$ which corresponds to a given PBH mass $M$ depends on the amplitude of the CDM fluctuation $\sigma_c(R,\itime)$ and therefore on the strength of the constraint on $\fPBH$, with grey lines showing curves of constant $M$. The requirement that $\fPBH < 1$ provides an isocurvature constraint across all scales (dashed) which weakens slightly slower than linearly in $\kR/\keq$ (see Eq.~\eqref{eq:sigmac_powerlaw}). The formation time of the PBHs is roughly $\aNL \simeq \sigma_c(R,\itime)^{-1}$, and the constraint on $\sigma_c(R,\itime)$ is roughly equivalent to the constraint the primordial isocurvature $\Delta_S(\kR)$ for power-law isocurvature power spectra through Eq.~\eqref{eq:sigmaR_powerlaw}. For further discussion see \S\ref{sec:constraints}.}
	\label{fig:bounds}
	\end{figure*}

	We show our matching procedure in Fig.~\ref{fig:backtracking} for three modes of a fixed scale $\kR/\keq = 10^6$. We highlight the small $\delta_c(R,0) \ll 1$ regime (red), the intermediate $\delta_c(R,0) \gg 1$ regime (blue) , and the limiting regime $\delta_c(R, \itime) \simeq \sqrt{2} \kR/\keq$ (purple) regimes. Each mode is matched with a perturbation of a matter only universe and
	PBH formation from the two modes is identified. This identification is a conjecture, and though its limiting behaviors are well understood analytically, future work and in particular simulations of PBH formation from CDM isocurvature fluctuations will be essential to improving the constraints we present here.

	Once they have formed, the abundance of PBHs relative to the CDM is related to their abundance at formation through a growth function $g(R)$,
	\begin{equation}
	\fPBH(R) \equiv \frac{\Omega_\PBH(R)}{\Omega_{c}} = g(R) \beta(R),
	\end{equation}
	where the growth function accounts for the relative redshifting of matter and radiation and therefore depends on when the PBHs form. The total abundance today is $f_\PBH = \int d \ln R\ f_\PBH(R)$. We associate the formation time of the black holes with the non-linearity time $\aNL$, and thus estimate the growth function as
	\begin{equation}
	\label{eq:growth}
	g(R) \sim \left( 1 + \frac{\aeq}{ \aNL}\right) \sim \frac{\sigma_c(R,0)}{\xi} ,
	\end{equation}
	where in the last approximation we work in the $\aNL \ll \aeq$ limit and we introduce an order unity fudge factor $\xi$ to account for uncertainty in the formation time. In our figures we take $\xi = 1$.

	We can now compute the primordial black hole abundance today resulting from Gaussian initial CDM isocurvature fluctuations. Since the CDM isocurvature is well-constrained on large scales \cite{Akrami:2018odb}, we are predominantly interested in small-scale perturbations $\kR \gg \keq$. In this regime $\sigmaHC$ is suppressed by $(\keq / \kR)^2$, and therefore achieving a sizable abundance of black holes requires a very large $\sigma_c(R,\itime)\gg1$. In this regime the auxiliary function $B \rightarrow 1$ and $\sigmaHC$ approaches 
	\begin{equation}
	\label{eq:sigmaHClarge}
	\sigmaHC \simeq  {\frac{\sqrt{3}}{2}} \frac{\keq^2}{\kR^2} \sigma_c(R,\itime)^2. 
	\end{equation}
	Since the abundance at formation $\beta = \fMD(\sigmaHC)$ is a steep function, significant PBH formation generally requires $\sigma_c(R,\itime) \sim \mathcal{O}(\kR/\keq)$.

	We show in Fig.~\ref{fig:betaforsigma} the PBH abundance, today and at formation, as a function of formation scale $\kR/\keq$ for a scale-dependent primordial isocurvature amplitude $\sigma_c(R,\itime) = 0.2 \kR / \keq$. For $\kR \gg \keq$, $\sigmaHC$ approaches a constant and thus so does the abundance at formation $\beta(R)$. The abundance today $\fPBH(R)$, however, grows as the increasing $\sigma_c(R,\itime)$ translates to a larger and larger growth function $g(R)\simeq \sigma_c(R,\itime)/\xi$ \eqref{eq:growth}.

\section{Isocurvature constraints from Primordial Black Holes}
\label{sec:constraints}

	The technique developed in \S\ref{sec:collapse} allows the present day PBH abundance to be computed for given initial CDM isocurvature fluctuations. We can now invert this procedure and map observational constraints on PBH abundances to constraints on the amplitude of primordial isocurvature. We assume in obtaining these constraints that the CDM isocurvature fluctuations are Gaussian\footnote{As discussed in Ref.~\cite{Domenech:2021and}, the isocurvature perturbation must by definition satisfy a positive-energy bound $S(0)(\vec{x}) \geq -1$. It must also have zero mean. To allow for regions of large $S(0)(\vec{x}) \gg 1$, most of the universe should therefore have $S(0)(\vec{x})=-1$ and be devoid of CDM. While this highly non-Gaussian distribution does not impact our  local perturbation matching procedure, it can effect the abundance of large $S$ regions relative to a Gaussian. Whether this change enhances or suppresses the PBH abundance depends on the specific distribution for $S$, and therefore we present here Gaussian results as simply a first estimate.}. The initial smoothed CDM variance $\sigma_c(R,\itime)^2$ in terms of the final PBH abundance $\fPBH(R)$ is, in the $\sigma_c(R,\itime) \gg 1$ and $\kR\gg\keq$ limits,
	\begin{equation}
	\label{eq:sigmac_general}
	\sigma_c(R,\itime)^2 \simeq {\frac{2}{\sqrt{3}}} \frac{\kR^2}{\keq^2} \fMD^{-1} \left(\frac{\fPBH(R)}{g(R)}\right),
	\end{equation}
	where $\fMD^{-1}$ is the inverse function of the PBH abundance for adiabatic fluctuations in matter domination $\fMD$. 	
	\begin{figure*}[t!]
	\includegraphics{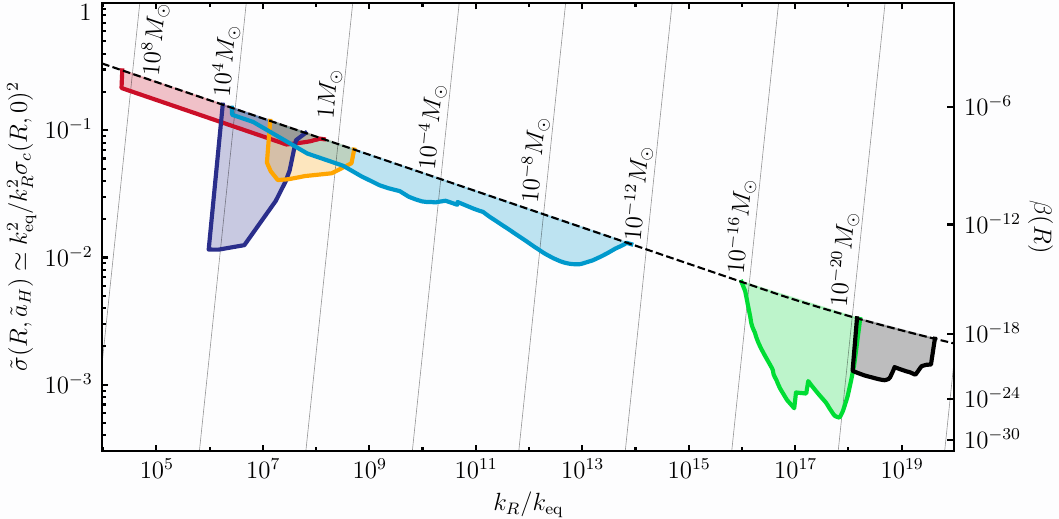}
	\caption{The value of the matched amplitude $\sigmaHC$ (left axis) of fluctuations of a matter-only universe which we use to compute the PBH abundance at formation $\beta(R)$ (right axis) through $\fMD$ (Eqs.~\eqref{eq:fharada} and \eqref{eq:fharada2}), as constrained by the PBH abundance today $\fPBH$ (Fig.~\ref{fig:bounds}). This figure is equivalent to the lower panel of Fig.~\ref{fig:bounds} with a $\kR/\keq$ scaling factored out since $\sigmaHC \simeq \keq^2/\kR^2 \sigma_c(R,0)^2$ (Eq.~\eqref{eq:sigmaHClarge}). At higher $\kR/\keq$, the PBHs form at earlier times in radiation domination and therefore a smaller value of $\beta$ can yield the same present day $\fPBH$. For $\sigmaHC\gtrsim5\times10^{-3}$, $\beta$ is given by \eqref{eq:fharada}. For $\sigmaHC\lesssim5\times10^{-3}$, $\beta$ becomes exponentially suppressed through \eqref{eq:fharada2}.  For further discussion see \S\ref{sec:constraints}.}
	\label{fig:sigmaHC}
	\end{figure*}

	Measurements of the PBH abundance are performed at a given mass as $\fPBH(M)$. Neglecting accretion, the mass $M$ of the PBH produced by a given isocurvature fluctuation obeys Eq.~\eqref{eq:MtoR}, depending not only on the scale $R$ of the perturbation but also on the amplitude of the perturbation $\sigma_c(R,\itime)$. The scale probed by measurements at a given mass $M$ is therefore 
	\begin{equation}
	\label{eq:kRofM_general}
	\frac{\kR(M)}{\keq} \simeq \left(\frac{\Meq}{M} \sigma_c(R,\itime)\right)^{1/3},
	\end{equation}
	again in the  $\sigma_c(R,\itime) \gg 1$ regime and where $\Meq \equiv M(\gamma/\keq)|_{\sigma_c = 0}\sim 10^{17} h^{-1} \Msun$ is roughly the horizon mass at matter-radiation equality. Plugging $\kR(M)$ into Eq.~\eqref{eq:sigmac_general} and reorganizing for the rms fluctuation yields
	\begin{equation}
	\sigma_c(R,\itime) \simeq  {\frac{2^{3/4}}{3^{3/8}}} \sqrt{\frac{\Meq}{M}} \fMD^{-1}  \left(\frac{\fPBH}{\sigma_c(R,\itime)/\xi}\right)^{3/4},
	\end{equation}
	where we have also replaced the growth function $g(R)$ with its form in the  $\sigma_c(R,\itime) \gg 1$ regime,  $\sigma_c(R,\itime)/\xi$. For a power-law $\fMD(x) =  b x^q$, with $\{b, q\}$ constants,
	we find
	\begin{equation}
	\label{eq:sigmac_powerlaw}
	\sigma_c(R,\itime) \simeq \left({\frac{2^{3/4}}{3^{3/8}}} \sqrt{\frac{\Meq}{M}}  \left(\frac{\xi \fPBH }{b}\right)^{\frac{3}{4 q}} \right)^{\frac{4 q}{3+4q}},
	\end{equation}
	which gives the amplitude of the constraint on the primordial isocurvature for a given measurement of $\fPBH(M)$. To find the specific scale that is being constrained we plug Eq.~\eqref{eq:sigmac_powerlaw} back into the $\kR(M)$ equation \eqref{eq:kRofM_general} to find
	\begin{equation}
	\label{eq:kRofM_powerlaw}
	\frac{\kR}{\keq} \simeq \left({\frac{2^{1/4}}{3^{1/8}}}  \left(\frac{\Meq}{M}\right)^{\frac{1}{2}+\frac{1}{4 q}} \left(\frac{ \xi \fPBH}{b}\right)^{\frac{1}{4q}} \right)^{\frac{4q}{3+4q}},
	\end{equation}
	which as expected depends not only on the mass $M$ but on the strength of the constraint $\fPBH(M)$. Finally, we can express the constraint on $\sigma_c(R,\itime)$ as a function of $\kR/\keq$ as
	\begin{equation}
	\label{eq:sigmaR_powerlaw_k}
	\sigma_c(R,\itime) \simeq \left({\frac{2^{1/2}}{3^{1/4}}} \frac{\kR}{\keq} \left(\frac{\xi\fPBH}{b}\right)^{\frac{1}{2q}} \right)^{\frac{2q}{2q+1}},
	\end{equation}
	where we see that for fixed $\fPBH$ the constraint on $\sigma_c(R,\itime)$ degrades slightly slower than linearly in $\kR/\keq$.

	We show in the top panel of Fig.~\ref{fig:bounds} a representative selection of constraints on the abundance of black holes of various masses, $\fPBH(M)$,  obtained with the help of the {\texttt{PBHbounds}} package \cite{Kavanagh:2019aaa}. As a whole, these constraints extend from $10^{-24}\Msun$ to $10^9 \Msun$ and at their strongest constrain PBHs to comprise less than $10^{-10}$ of the CDM density. The constraints we use are from:
	\begin{itemize}
	\item The effect of PBH evaporation on big bang nucleosynthesis (BBN) \cite{Carr:2009jm,Acharya:2020jbv};
	\item The effect of PBH evaporation on the CMB \cite{Clark:2016nst, Acharya:2020jbv}, EDGES \cite{Clark:2018ghm,Mittal:2021egv}, INTEGRAL \cite{Laha:2020ivk}, the 511 keV line \cite{Laha:2019ssq,DeRocco:2019fjq}, Voyager \cite{Boudaud:2018hqb}, extragalactic gamma rays \cite{Carr:2009jm}, and the interstellar medium \cite{Laha:2020vhg}; 
	\item Microlensing by PBHs of EROS \cite{EROS-2:2006ryy},  HSC \cite{Niikura:2017zjd,Smyth:2019whb,Croon:2020ouk}, Kepler \cite{Griest:2013aaa} and MACHO \cite{Macho:2000nvd} objects;
	\item The number of coalescences detected by LIGO-Virgo \cite{Kavanagh:2018ggo,LIGOScientific:2018mvr};
	\item Accretion onto PBHs altering the reionization history \cite{Serpico:2020ehh};
	\item Dynamical disruption of wide binaries by PBHs \cite{2014ApJ...790..159M}.
	\end{itemize}
	Additional constraints can be found compiled in, e.g., Refs.~\cite{Villanueva-Domingo:2021spv}~and~\cite{Carr:2020gox}. The constraints shown here are not necessarily the strongest across their respective mass ranges (compare, e.g., those from PBH gas heating \cite{Takhistov:2021aqx}). 

	In the bottom panel of Fig.~\ref{fig:bounds}, we map these PBH constraints into constraints on the smoothed primordial CDM isocurvature amplitude $\sigma_c(R, \itime)$, using Eq.~\eqref{eq:sigmac_general} and Eq.~\eqref{eq:kRofM_general} with the full $\fMD$ \eqref{eq:fharada} including the exponential suppression from Eq.~\eqref{eq:fharada2}. Since constraints at given mass $M$ map to different primordial scales $\kR$ depending on the amplitude of the constraint $\fPBH$, we show lines of fixed PBH mass $M$ to guide the eye. The constraint $\fPBH < 1$, dashed, holds for all PBHs which have an evaporation time longer than the age of the universe at BBN, corresponding to masses $M \gtrsim 10^{-24} \Msun$. For constant $\fPBH(M)$, the constraint on $\sigma_c(R,\itime)$ weakens slightly slower than linearly in $\kR/\keq$, as derived in Eq.~\eqref{eq:sigmaR_powerlaw_k}. The formation time of the PBHs $\xi \aNL \sim \xi/\sigma_c(R,\itime)$ can also be read from this figure. 

	We factor out the $\kR/\keq$ scaling of our constraint on $\sigma_c(R,0)$ by showing in Fig.~\ref{fig:sigmaHC} the intermediate matching parameter $\sigmaHC\simeq \keq^2/\kR^2 \sigma_c(R, \itime)^2$ (see Eq.~\eqref{eq:sigmaHClarge}) that goes into our computation and which directly determines the PBH abundance at formation $\beta(R)$ through Eqs.~\eqref{eq:fharada} and \eqref{eq:fharada2}. For larger values of $\kR/\keq$, the PBHs form earlier in radiation domination and therefore the same abundance today can be achieved with smaller $\beta$. The exponential suppression of PBH formation \eqref{eq:fharada2} in the small $\sigmaHC$ limit is reflected in the compression of logarithmically spaced tick-marks in $\beta(\sigmaHC)$.

	Finally, constraints on $\sigma_c(R, \itime)$ can be converted to constraints on the primordial isocurvature spectrum $\Delta^2_S (k)$ through
    \begin{equation}
	\sigma_c^2 (R, \itime) = \int \frac{d k}{k} W_R(k)^2  \Delta^2_S(k).
	\end{equation}
	In principle the primordial power spectrum in Fourier space can be reconstructed model independently if $\sigma_c^2(R,\itime)$ is known across a wide range of scales \cite{Kimura:2021sqz}, but results for specific choice of $W_R(k)$ and $\Delta^2_s(k)$ are illuminating.	For example, for a power-law isocurvature spectrum $\Delta^2_S =  A \left(k/k_*\right)^{n_{\rm iso}-1}$ and a Fourier space top hat window function, we have
	\begin{equation}
	\label{eq:sigmaR_powerlaw}
	\sigma_c^2 (R,\itime) =  \Delta_S^2(\kR) c_1,
	\end{equation}
	with $c_1 \equiv \left(1-(\kmin /\kR)^{n_{\rm iso}-1}\right)/(n_{\rm iso} - 1)$ a number which diverges as an infrared cutoff $\kmin\rightarrow0$ for $n_{\rm iso} \leq 1$ but otherwise is generally of order unity. The constraint on smooth, power-law isocurvature spectra can therefore be obtained directly from Fig.~\ref{fig:bounds}.

	For an isocurvature power spectrum characterized instead by a log-normal peak at $k_P$ with width $d$ and amplitude $A(k_p)^2$ with functional form 
	\begin{equation}
	\Delta_S^2(k) = A(k_p)^2 \frac{1}{\sqrt{2 \pi} d} \exp\left[-\frac{\log^2(k/k_p)}{2 d^2}\right],
	\end{equation} 
	the variance of the smoothed density contrast is
	\begin{equation}
	\label{eq:sigmaR_peak}
	\sigma_c^2 (R,\itime) =  A(k_p)^2 \frac{\erfc}{2}\left[\frac{1}{\sqrt{2} d} \ln\left[\frac{k_p}{\kR}\right]\right],
	\end{equation}
	where $\erfc(x) = 1 - \erf(x)$ is the complimentary error function, which has limits $\erfc(x \ll0) \rightarrow  2$,  $\erfc(x\gg0) \rightarrow 0$. In the narrow peak limit $d \rightarrow 0$, constraints on $\sigma_c(R, \itime)$ therefore place an upper bound on $A(k_P)^2$ on all large scales $k_P < \kR$,
	\begin{equation}
	\label{eq:Apbound}
	A(k_p)^2 \lesssim \min(\sigma_c(\kR > k_P, \itime)^2).
	\end{equation}

	\begin{figure}[t!]
	\includegraphics{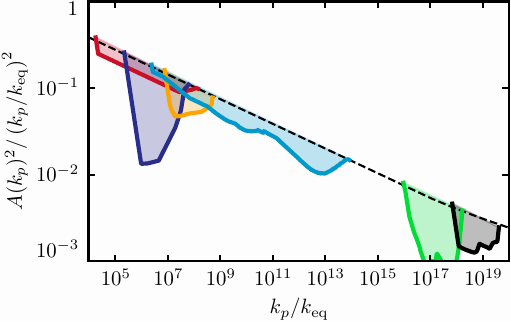}
	\caption{The constraint from PBH abundances on the amplitude $A(k_p)^2$ of a log-normal peak at $k_p$ in the isocurvature power spectrum $\Delta^2_S$ in the narrow peak limit. The constraint on $A(k_p)^2$ degrades slightly slower than quadratically. For further discussion see \S\ref{sec:constraints} and Eq.~\eqref{eq:Apbound}.}
	\label{fig:ap}
	\end{figure}

	We show this constraint on the amplitude of a narrow log-normal peak in the isocurvature power spectrum in Fig.~\ref{fig:ap}. Since the power spectrum scales as the square of the perturbation, the constraint on the power spectrum amplitude weakens slightly slower than quadratically.

\section{Discussion}
\label{sec:discussion}

	We have a presented a new mechanism to form PBHs: the collapse of large primordial CDM isocurvature fluctuations. We have used this mechanism to provide  constraints on the amplitude of isocurvature perturbations between $\sim 10^3 \ \Mpc^{-1}$ and $10^{17} \ \Mpc^{-1}$ which are independent of the dark-matter model. Our constraints are much weaker than those on the adiabatic mode, but we are aware of no {\it a priori} reason why the CDM isocurvature mode could not be so large.

	Large isocurvature perturbations on small scales can be formed from, e.g., an axion field \cite{Kasuya:2009up} or the formation of compact objects \cite{Gong:2017sie,Gong:2018sos}, but we are not aware of any model which produces isocurvature perturbations of this size. {Our constraints assume the isocurvature fluctuations are Gaussian, but any realistic model which produces sufficiently large isocurvature must produce a highly skewed non-Gaussian distribution which may impact the resulting PBH abundance.}	{We neglected baryons throughout our calculation, an excellent approximation for PBHs formed on very small scales deep in radiation domination since the baryons are then energetically insignificant.}

	That PBHs can form from isocurvature perturbations of a matter-like fluid if they are so large that they lead to a significant total density perturbation as they enter the horizon was perhaps first noted in Ref.~\cite{Yokoyama:1995ex}, and such a setup has been used to produce PBHs at the beginning of an early matter-dominated epoch (see, e.g., \cite{Cotner:2016cvr,Cotner:2018vug,Cotner:2019ykd}). Our results are the first to make use of just primordial CDM fluctuations evolving during a traditional cosmic history.

	Our constraints extend to much smaller scales than existing model-independent constraints. The CMB severely constrains primordial CDM isocurvature up to scales $k\lesssim 0.1 \ \Mpc^{-1}$ \cite{Akrami:2018odb}, and large scale structure and Lyman-$\alpha$ data extend the constraints to megaparsec scales \cite{Seljak:2006bg,Croft:1997jf,Sugiyama:2003tc,Beltran:2005gr}. Future 21cm constraints may reach $k\sim 1000 \ \Mpc^{-1}$ \cite{Sekiguchi:2013lma,Furugori:2020jqn}, and $\mu$-distortions could in principle constrain up to $k\sim 10^4 \ \Mpc^{-1}$ \cite{Chluba:2013dna}.

	Much more powerful constraints are possible assuming specific dark matter models. If the dark matter annihilates \cite{Kohri:2014lza,Nakama:2017qac} or decays \cite{Yang:2014lsg}, for example, strong isocurvature constraints can be placed from $\gamma$-ray and neutrino emission. And if the dark matter is comprised of thermally produced WIMPs, kinetic decoupling sets a cutoff in the small scale CDM power spectrum \cite{Bringmann:2009vf}.

	An open question is the relationship of our results to constraints on dark matter substructure \cite{Ricotti:2009bs,Blinov:2021axd,Li:2012qha,Delos:2017thv}, which is the subject of significant recent interest and can be constrained by, e.g., caustic microlensing \cite{Diego:2017drh,Oguri:2017ock,Dai:2019lud} or pulsar timing \cite{Siegel:2007fz,Clark:2015sha,Dror:2019twh,Ramani:2020hdo,Lee:2020wfn}, but depends sensitively on the nature of the dark matter and its nonlinear evolution. On the same note, the details of our results rely on an conjecture for the probability of CDM fluctuations to collapse to PBHs, which deserves further validation from numerical simulations. 

	As future constraints on PBH abundances arrive they can be converted to improved CDM isocurvature constraints using our results. However, because of the exponential suppression of PBH formation in the small $\sigmaHC$ limit, the constraints presented here can probably not be improved dramatically, especially on the smallest scales. The PBH constraints from evaporation, for example, approach the exponential floor where future improvements to $\fPBH$ can lead to only logarithmic improvements in the isocurvature constraint. This is similar to PBH constraints on small-scale adiabatic perturbations, where the abundance is always exponential in the perturbation amplitude. 

	Instead, a key signal of large fluctuations on small scales is the production of second-order induced gravitational waves (see, e.g., \cite{Domenech:2021ztg} for a recent review). Constraints on induced gravitational waves from future gravitational wave detectors will lead to constraints on the small-scale adiabatic spectrum stronger than the constraints from PBH production. {Following up on this work, the induced gravitational wave signal from small-scale CDM isocurvature perturbations which collapse during global radiation domination is studied in Ref.~\cite{Domenech:2021and}.}

\acknowledgments
We are grateful to Metin Ata, Guillem Dom\`{e}nech, Wayne Hu, Kazunori Kohri, Volodymyr Takhistov, and Valeri Vardanyan for insightful comments and discussions. This work was made possible by the Yukawa Institute for Theoretical Physics workshop YITP-X-21-02 and by the World Premier International Research Center Initiative (WPI), MEXT, Japan. This work was supported in part by JSPS KAKENHI Nos.~19H01895, 20H04727, and 20H05853.

\bibliographystyle{apsrev4-1}
\bibliography{references.bib}

\end{document}